# Achieving Dependability of AI Execution with Radiation-Hardened Processors

Carlos Rafael Tordoya T., Hans Dermot Doran
Institute of Embedded Systems
ZHAW School of Engineering
Winterthur, Switzerland
tord@zhaw.ch, donn@zhaw.ch

Pablo Ghiglino, Mandar Harshe
Klepsydra Technologies AG
Zug, Switzerland
pablo.ghiglino@klepsydra.com,
mandar.harshe@klepsydra.com

*Abstract*—**The reliance on radiation-hardened hardware, essential for domains requiring high-dependability such as space, nuclear energy and medical applications, severely restricts the choice of components available for modern AI-intensive tasks, particularly for real-time AI-based classifications. To address this challenge, we propose leveraging the High Performance Data Processor (HPDP) as a radiation-hardened and low-power co-processor in conjunction with an optimized AI framework for efficient data processing.**

**The HPDP's dynamic reconfiguration capabilities and dataflow-oriented architecture provide an ideal platform for executing AI-driven applications that demand low-latency, high-throughput streaming data processing. To fully utilize the co-processor's capabilities, we utilized Klepsydra's AI-runtime inference framework, which, due to its lock-free execution and efficient resource management, significantly enhances data processing throughput without increasing power consumption.**

**Our approach entails programming the HPDP as a dedicated mathematical backend, enabling the AI framework to execute workloads directly on this co-processor without requiring additional hardware-specific coding. This paper presents the preliminary results of our implementation, describing the application domain, AI pipeline, key features of the HPDP architecture, and performance evaluation. Our solution demonstrates a significant advancement in deploying AI on radiation-hardened platforms by using the HPDP as a dependable, efficient, and reprogrammable co-processor, making it highly suitable for any application requiring dependable execution in any environment.**

## I. INTRODUCTION

### 1. Motivation

The deployment of Edge AI (artificial intelligence executed directly on edge devices) in domains requiring high dependability, such as space exploration, nuclear operations, and medical diagnostics, has become increasingly essential [1], [2]. By processing data locally, Edge AI enables real-time decision-making, reduces processing latency, and minimizes reliance on cloud-based resources, which is critical in domains with limited connectivity. However, harsh environmental conditions, particularly radiation exposure, pose significant challenges to electronic systems in these domains. Radiation can disrupt electronic circuits, resulting in data corruption, operational disruptions, and compromised system reliability [3].

To maintain reliable operation under such conditions, redundant execution or radiation-hardened processors are commonly utilized. Radiation-hardened processors offer robust resistance to radiation, ensuring reliable performance in extreme environments. However, their market longevity and computational capacity tends to be limited compared to frequently upgraded commercial processors, questioning their ability to handle computationally intensive workloads such as convolutional neural networks (CNNs). Therefore, attempts to integrate Edge AI into radiation-hardened systems often result in reduced performance compared to commercial solutions and/or the need for highly specialized programming [4] because they lack the computational power of commercial CPUs and GPUs. This highlights the need for innovative approaches to bridge the gap between the dependability of radiation-hardened hardware and the computational demands of Edge AI.

## 2. Proposed solution

This paper proposes leveraging the High-Performance Data Processor (HPDP) as a radiation-hardened co-processor to overcome these challenges. The HPDP's dynamic reconfiguration capabilities and parallel, dataflow-oriented architecture provide an optimized platform for running computationally intensive tasks, such as convolution operation in CNN inference, with low latency and high throughput.

To enable a complete AI execution system, the RTG4 from Microchip Technology [5] (a radiation-tolerant FPGA with an integrated LEON3 processor) acts as the main orchestrator for HPDP operations. It manages data transfers to and from the HPDP, triggers AI workloads on this co-processor, and receives orchestration instructions from a payload computer running the Klepsydra AI framework (see Figure 1). By offloading computational tasks to the HPDP, this approach combines computational efficiency with system dependability, which is essential for applications in radiation-prone environments.

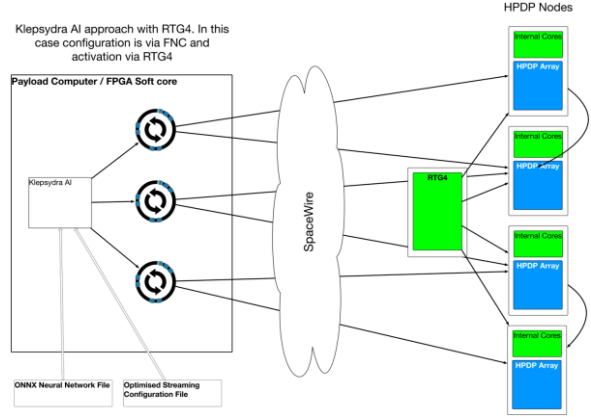

Figure 1. HPDP for AI integration diagram

## II. RELATED WORK

Artificial Intelligence (AI) deployment in radiation-exposed conditions faces significant challenges due to the computational limitations of radiation-hardened processors. Widely used processors, such as the RAD750 [6] and LEON [7] series, are reliable under radiation exposure but trail significantly behind commercial processors in terms of computational density and power efficiency. For example, the computational density of radiation-hardened processors, including the RAD750 and GR740 (LEON4), is significantly lower than commercial processors like the ARM Cortex-A9, commonly found in systems such as the Xilinx Zynq 7020 [8]. These performance limitations severely restrict the size and complexity of AI models, making the deployment of modern AI workloads like Convolutional Neural Networks (CNNs) generally infeasible on such platforms. As a result, bridging the gap between the dependability of radiation-hardened hardware and the computational demand of AI remains a critical challenge [9].

Reconfigurable architectures, such as Coarse-Grained Reconfigurable Arrays (CGRAs), with their coarse-grained processing elements like ALUs and small processing units, offer flexibility and parallelism for high-performance workloads. However, they are not inherently radiation-hardened, making them susceptible to faults like bit flips and single-event upsets (SEUs) during their operation in radiation-prone environments. Even with complex fault-tolerant mechanisms, the overhead required to maintain reliability tends to limit their applicability in such environments [10]. The High-Performance Data Processor, developed by the European Space Agency (ESA) and Airbus, addresses these challenges with its dynamically reconfigurable architecture. Supporting high-throughput and parallel data processing at a clock speed of 250 MHz [11], the HPDP has been applied in tasks such as FFT and data compression [12].

## III. HPDP ARCHITECTURE

### 1. Core architecture

The High-Performance Data Processor is built around the XPP-III core (XPP array); a dynamically reconfigurable processing array optimized for parallelism. This core consists of 40 Arithmetic Logic Unit Processing Array Elements (ALU-PAEs), arranged in a 5x8 matrix, and 16 RAM-PAEs, organized in 2x8 columns on either side of the ALU array [13] (see Figure 2). These PAEs are designed to handle a wide range of computational and data storage tasks efficiently.

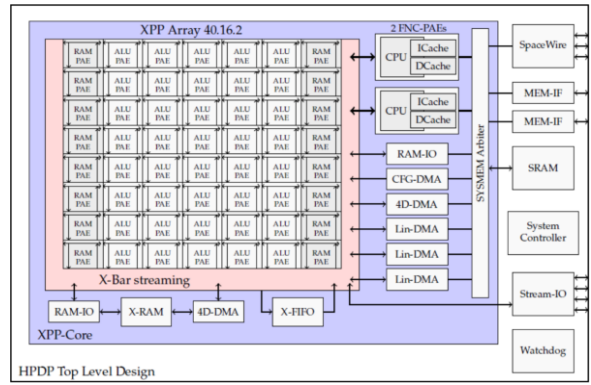

Figure 2. HPDP architecture

The HPDP integrates robust memory and I/O interfaces to support high-throughput data transfer:

- 4D-DMA: Enable data transfers for HPDP memory with complex address patterns.
- SpaceWire Interface: Facilitates communication with external systems.
- Stream-IO interfaces: Enables data streaming to and from the XPP array.

The FNC-PAEs manage memory interfaces, DMA transfers, XPP array reconfiguration, sequential execution, and data exchange with the XPP array [13].

### 2. XPP array dataflow

Each PAE operates independently and in parallel. This architecture allows data to be processed as a continuous stream that flows through the XPP array PAEs while being processed [12]. With a clock speed of up to 250 MHz, the XPP array is capable of providing optimized throughput and low-latency performance [11], which makes it suitable for data-intensive applications that can benefit from parallel execution.

### 3. HPDP configuration

To configure the HPDP, two main components should be programmed [14]:

- FNC-PAE: programmed in C.
- XPP array: Programmed using NML (Native Mapping Language) code, which is a programming language for describing configurations on the XPP Dataflow Array.

Code can run on the FNC-PAE without programming the XPP array, but not vice versa.

In addition, a vectorizing C Compiler called XPP-VC [15] can be used to configure both FNC-PAE and XPP array. This compiler partitions a single C program into C code for FNC-PAE and C code for XPP array, converting the latter into NML code.

## IV. HPDP AS A DEPENDABLE CO-PROCESSOR FOR AI EXECUTION

### 1. Implementation Strategy

In this study the High-Performance Data Processor (HPDP) is employed as a co-processor to execute Edge AI workloads. To achieve this, the XPP array was programmed using NML language to execute computationally intensive operations, particularly convolutions. Additionally, the FNC-PAE was programmed using C language to configure the DMA modules, load the XPP configuration, and setup interfaces for data transfer between XPP array and HPDP memory.

The implementation process of the convolution operation in the XPP array followed an iterative process of validation and refinement. Initially, simple multiplication and addition operations were tested on the PAEs to ensure correct functionality. Then, a basic convolution was implemented to analyze dataflow behavior, verifying that data streaming was handled correctly. Based on these insights, the implementation was progressively improved, resulting in a design where convolution and re-quantization [16] operations in XPP array rely solely on input parameters (weights, bias, input activation and re-quantization parameters) to execute. In addition, these two operations process the data stream in parallel, ensuring continuous execution without introducing additional delays.

### 2. Verification Methodology

During the development process of the convolution and re-quantization operations in the XPP array, verification was performed by using the XPP debugger XDBG [15] in the following ways:

- This tool was used to display the XPP array configuration, which enabled verification of whether all the PAEs were interconnected properly and recognition of potential bottlenecks in the routing operations of the input data stream.
- This tool also enabled detailed monitoring of the operations performed by each PAE and allowed visualization of how the data stream flows through the XPP array at each simulation cycle (see Figure 3). This made it possible to verify which PAEs were working as expected and to check intermediate results throughout the entire process of convolution and re-quantization.

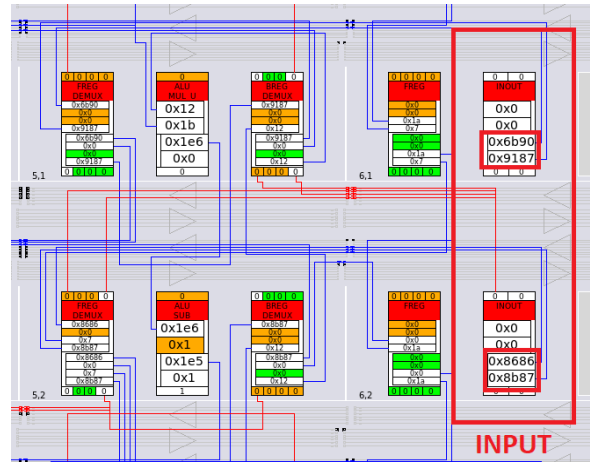

Figure 3. PAEs visualization example using XDBG tool

### 3. Validation Methodology

As part of the iterative refinement process, a validation framework was developed to ensure the correctness of the convolution implementation on HPDP. This framework compared HPDP's output against a PyTorch-based [17] reference implementation and was executed within a unit-test based testing environment to verify correctness and performance (see Figure 4).

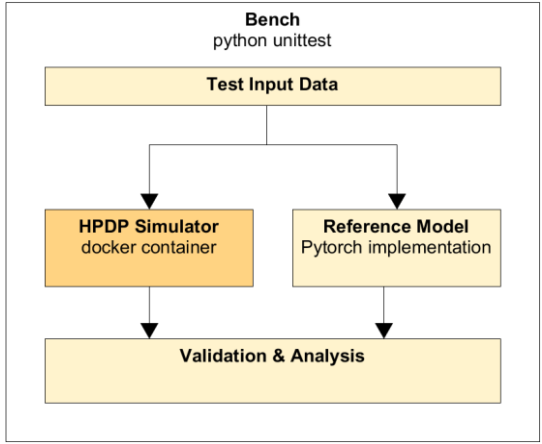

Figure 4. Validation methodology diagram

The HPDP simulator, based on the processor's datasheet, provides a cycle-accurate representation of execution behavior. This ensures that results obtained from the simulation closely reflect those expected on actual HPDP hardware. Therefore, the convolution algorithm was executed within this simulated environment, and the output feature maps were extracted for comparison.

A functionally equivalent convolution operation was implemented using PyTorch, which provided the expected output and enabled the detection of any deviations in HPDP's output. In cases where differences were observed, further analysis was conducted to identify potential numerical differences in intermediate computations or execution inconsistencies.

To systematically validate the HPDP convolution results, the unit-test framework was used for structured testing. The validation procedure included:

- Numerical comparison of HPDP's output feature maps against PyTorch's reference results to check for deviations.
- Performance monitoring, generating plots that visualize data processing rates across different convolution implementations.

This validation framework ensured that HPDP's convolution and re-quantization produced correct and consistent results.

*4. System Integration and Execution Flow*

In the final solution, the Klepsydra AI framework runs on a payload computer, providing high-level orchestration for AI workloads based on two inputs: an ONNX-based neural network model [18] and an Optimized Streaming Configuration file [19] provided by Klepsydra AI. The framework sends orchestration instructions to RTG4, which acts as the main orchestrator for HPDP operations. RTG4 manages data transfers via Stream-IO, configures the XPP array and DMA modules through FNC, and triggers the convolution process on HPDP (see Figure 1).

During execution, HPDP autonomously performs convolution and re-quantization operations, storing the processed data in its internal memory. The RTG4 then controls where the output data are sent:

- Back to the RTG4 via Stream-IO for further processing.
- Directly to another HPDP, allowing it to immediately process the next AI layer without additional data transfer.

This dataflow enables efficient AI processing, with RTG4 ensuring smooth execution flow and HPDP providing parallel, high-throughput computation.

V. RESULTS

Following the iterative development and refinement process, the convolution algorithm for the High-Performance Data Processor (HPDP) reached a stable implementation, enabling benchmarking tests. These tests aimed to evaluate the computational performance of HPDP by comparing its convolution processing latency with the results obtained from the GR740, a radiation-hardened processor featuring a LEON4 (SPARC V8) architecture running at a 250 MHz clock speed.

In order to align the benchmarking tests with a real use case, the Ship Detection application within the OBPMark-ML suite (set of benchmarks that cover applications commonly found in on-board spacecraft) [20] was selected. In this application, satellites capture images from Earth's surface that then are processed using CNN models, like YoloX, to identify and track ships.

Therefore, the benchmarking process measured the processing time required to execute convolution and re-quantization operations for specific layers of a CNN model for Ship Detection. Since the HPDP is currently implemented to perform only the convolution operation (with re-quantization) and relies on the RTG4 for overall CNN orchestration, the tests focused on individual convolution layers. To maintain consistency, all tested processors followed the same approach, evaluating only the quantized convolution and re-quantization operations of single layers.

In addition, to ensure a consistent evaluation, the input activation data, kernel, bias, and re-quantization parameters were preloaded into memory for all tested processors. This approach eliminated the influence of data transfer times from external devices (e.g., RTG4 to HPDP), ensuring that the measured latency reflected only the processing performance.

The preliminary results, presented in Table 1 and Figure 5, show the processing latency (in milliseconds) for different convolution layers, with kernel sizes indicating the specific layers tested.

| Model | Kernel Size | Image Size | HPDP | GR740 |
|---|---|---|---|---|
| Ship Detection Quantised | 24x3x3x24 | 194x194x24 | 121.27 ms | 23894.08 ms |
| | 48x3x3x48 | 98x98x48 | 110.94 ms | 23731.64 ms |
| | 96x3x3x96 | 50x50x96 | 104.84 ms | 11765.59 ms |
| | 96x1x1x96 | 96x96x96 | 47.44 ms | 31320.04 ms |

Table 1. Processing latency benchmarking results

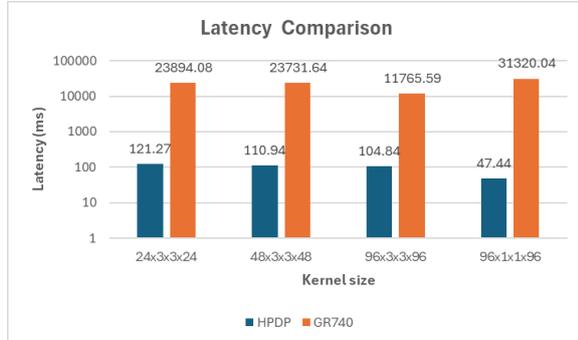

Figure 5. Latency comparison across processors

## VI. CONCLUSIONS

The benchmarking results demonstrated that HPDP consistently achieved competitive performance, particularly among radiation-hardened processors, with lower execution times than GR740 for all tested convolution layers, highlighting its computational efficiency.

Beyond its performance metrics, the HPDP advantage is its radiation-hardened design, essential for reliable operation in challenging environments, such as space missions, nuclear energy systems, and medical devices. Additionally, HPDP's dataflow-oriented architecture aligns seamlessly with the Klepsydra AI-runtime inference framework, facilitating efficient integration into streaming-based AI systems without requiring additional hardware-specific programming once configured.

This combination of computational efficiency, dependability, and compatibility with dataflow-driven AI frameworks establishes HPDP as a strong candidate for highly dependable applications where radiation resistance and low-latency AI processing are critical requirements.

## VII. FUTURE WORK

Future work will focus on expanding benchmarking tests to include additional radiation-hardened and commercial processors, which will provide a broader comparison of the HPDP's computational performance. Additionally, future work will involve considering more performance metrics to gain a deeper understanding of the HPDP's suitability for AI execution in radiation-prone environments, as well as exploring advanced memory management techniques to optimize data transfer efficiency between XPP array and HPDP memory to enhance overall performance.

## REFERENCES

[1] A. Russo and G. Lax, "Using Artificial Intelligence for Space Challenges: A Survey," Appl. Sci., vol. 12, no. 10, p. 5106, May 2022. doi: 10.3390/app12105106.

[2] Y. Xu, T. M. Khan, Y. Song, and E. Meijering, "Edge deep learning in computer vision and medical diagnostics: a comprehensive survey," Artif. Intell. Rev., vol. 58, p. 93, 2025. doi: 10.1007/s10462-024-11033-5.

[3] [1] Y. Q. de Aguiar, F. Wrobel, J.-L. Autran, and R. García Alía, "Radiation Environment and Their Effects on Electronics," in *Single-Event Effects, from Space to Accelerator Environments*, Cham, Switzerland: Springer, 2025. doi: 10.1007/978-3-031-71723-9_1.

[4] G. Furano, G. Meoni, A. Dunne, D. Moloney, V. Ferlet-Cavrois, A. Tavoularis, J. Byrne, L. Buckley, M. Psarakis, K.-O. Voss, and L. Fanucci, "Towards the Use of Artificial Intelligence on the Edge in Space Systems: Challenges and Opportunities," *IEEE Aerosp. Electron. Syst. Mag.*, vol. 35, no. 12, pp. 44–56, Dec. 2020. doi: 10.1109/MAES.2020.3008468.

[5] Microchip Technology Inc., "RTG4 FPGA Datasheet". Accessed: Jan. 17, 2025 [Online]. Available: https://ww1.microchip.com/downloads/aemDocuments/documents/FPGA/ProductDocuments/DataSheets/RTG4_FPGA_Datasheet.pdf.

[6] R. W. Berger, D. Bayles, R. Brown, S. Doyle, A. Kazemzadeh, K. Knowles, D. Moser, J. Rodgers, B. Saari, D. Stanley, and B. Grant, "The RAD750™—A Radiation Hardened PowerPC™ Processor for High Performance Spaceborne Applications," in *2001 IEEE Aerospace Conference Proceedings* (Cat. No.01TH8542), Big Sky, MT, USA, Mar. 10-17, 2001, pp. 2263–2272. doi: 10.1109/AERO.2001.931184.

[7] J. Andersson, M. Hjorth, F. Johansson, and S. Habinc, "LEON Processor Devices for Space Missions: First 20 Years of LEON in Space," in *Proceedings of the 2017 IEEE 6th International Conference on Space Mission Challenges for Information Technology (SMC-IT)*, Pasadena, CA, USA, 2017, pp. 1–8. doi: 10.1109/SMC-IT.2017.31

[8] AMD, "Zynq-7000 Overview", AMD, Mar. 2025. Accessed: Jan. 20, 2025. [Online]. Available: https://docs.amd.com/v/u/en-US/ds190-Zynq-7000-Overview.

[9] J. Goodwill, C. Wilson, and J. MacKinnon, "Current Technology in Space," NASA Goddard Space Flight Center, Greenbelt, MD, USA, Jul. 2023. Accessed: Jan. 20, 2025 [Online]. Available: https://ntrs.nasa.gov/api/citations/20240001139/downloads/Current%20Technology%20in%20Space%20v4%20Briefing.pdf

[10] G. Lee, E. Cetin, and O. Diessel, "Fault Recovery Time Analysis for Coarse-Grained Reconfigurable Architectures," *ACM Trans. Embed. Comput. Syst.*, vol. 17, no. 2, Art. no. 42, pp. 1–21, Nov. 2017. doi: 10.1145/3140944.

[11] Airbus Defence and Space, "High Performance Data Processor (HPDP) Payload", Airbus, Dec. 2020. Accessed: Feb. 14, 2025. [Online]. Available: https://www.airbus.com/sites/g/files/jlcbta136/files/2021-11/publication-sce-payload-hpdp-12-2020.pdf

[12] M. A. Syed and E. Schueler, "High Performance Data Processor (HPDP)," 2008 NASA/ESA Conference on Adaptive Hardware and Systems. Noordwijk, Netherlands, 2008, pp. 178-182, doi: 10.1109/AHS.2008.45.

[13] G. Vives Vallduriola, T. Helfers, D. Bretz, M. Syed, D. Witsch, C. Papadas, V. Pérel, and S. Bartels, "High Performance Data Processor (HPDP) – Image Processing Applications of a New Generation Space Processor," in *ESA On-Board Payload Data Processing Workshop (OBDP 2019)*, Noordwijk, Netherlands, 2019.

[14] V. Baumgarte, G. Ehlers, F. May, A. Nückel, M. Vorbach, and M. Weinhardt, "PACT XPP—A self-reconfigurable data processing architecture," The Journal of Supercomputing, vol. 26, no. 2, pp. 167–184, 2003, doi: 10.1023/A:1024499601571.

[15] XPP Technologies, "Programming XPP-III Processors," White Paper, Version 2.0.1, Jul. 13, 2006. Accessed: Feb. 20, 2025. [Online]. Available: https://courses.cs.washington.edu/courses/cse591n/06au/papers/XPP-III_programming_WP.pdf

[16] B. Jacob, S. Kligys, B. Chen, M. Zhu, M. Tang, A. Howard, H. Adam, and D. Kalenichenko, "Quantization and Training of Neural Networks for Efficient Integer-Arithmetic-Only Inference," arXiv, Dec. 2017. [Online]. Available: https://arxiv.org/abs/1712.05877v1.

[17] M. C. Chirodea, O. C. Novac, C. M. Novac, N. Bizon, M. Oproescu, and C. E. Gordan, "Comparison of Tensorflow and PyTorch in Convolutional Neural Network - based Applications," in *2021 13th International Conference on Electronics, Computers and Artificial Intelligence (ECAI)*, Pitesti, Romania, Jul. 1-3, 2021. IEEE, doi: 10.1109/ECAI52376.2021.9515098

[18] V. Shankar, "Edge AI: A Comprehensive Survey of Technologies, Applications, and Challenges," in 2024 1st International Conference on Advanced Computing and Emerging Technologies (ACET), Ghaziabad, India, Aug. 2024, doi: 10.1109/ACET61898.2024.10730112.

[19] ESA, "ADCSS 2022 Presentation," in *16th ESA Workshop on Avionics, Data, Control, and Software Systems (ADCSS 2022)*, Noordwijk, Netherlands, Oct. 2022. [Online]. Available: https://indico.esa.int/event/421/contributions/6958/attachments/4925/7713/adcss_2022_presentation.pdf.

[20] D. Steenari, L. Kosmidis, I. Rodriguez-Ferrandez, A. Jover-Alvarez, and K. Förster, "OBPMark (On-Board Processing Benchmarks) – Open Source Computational Performance Benchmarks for Space Applications," in *2nd European Workshop on On-Board Data Processing (OBDP2021)*, Online, Jun. 14-17, 2021. Accessed: Feb. 21, 2025 [online] Available: https://zenodo.org/records/5638577